\documentclass[12pt]{article}
\usepackage{graphics}
\usepackage{amsmath,amsfonts}

\usepackage{amsmath}
\usepackage{amssymb}
\usepackage{amsthm}
\usepackage{psfrag}
\usepackage{graphicx}

\newcommand{\vev}[1]{\langle  #1 \, \rangle}

\newcommand{\barr}{}
\newcommand{\ma}{{\rm \mbox{{\scriptsize  M}}}}

\newcommand{\dil}{{\rm \mbox{{\scriptsize  dil}}}}
\newcommand{\eff}{{\rm \mbox{{\scriptsize  eff}}}}
\newcommand{\UV}{{\rm \mbox{{\tiny  UV }}}}
\newcommand{\IR}{{\rm \mbox{{\tiny  IR }}}}

\newcommand{\beq}{\begin{equation}}
\newcommand{\eeq}{\end{equation}}
\newcommand{\beqa}{\begin{eqnarray}}
\newcommand{\eeqa}{\end{eqnarray}}
\newcommand{\bea}{\begin{eqnarray}}
\newcommand{\eea}{\end{eqnarray}}
\newcommand{\LL}{{\cal L}}
\newcommand{\OO}{{\cal O}}

\newcommand\vp{\phi}
\newcommand\eepsilon{\Delta_-}

\title{
\sc{\huge Naturally light dilatons from\\  nearly marginal deformations
\\~}
\date{}
}

\author{
  Eugenio Meg\'ias\thanks{emegias@ifae.es} ~ and ~ Oriol Pujol\`as\thanks{pujolas@ifae.es}
\vspace{.5cm}\\
\normalsize\llap{}
 \it Departament de F\'isica and IFAE,
\normalsize\it Universitat Aut\` onoma de Barcelona,\\
\normalsize\it Bellatera 08193, Barcelona, Spain\\
\vspace{.5cm}\\
} 

\begin{document}
\maketitle
\begin{abstract}
We discuss the presence of a light dilaton in CFTs deformed by a nearly-marginal operator $\OO$, in the holographic realizations consisting of confining RG flows that end on a soft wall. Generically, the deformations induce a condensate $\vev\OO$, and  the dilaton mode can be identified as the fluctuation of $\vev\OO$. We obtain a mass formula for the dilaton as a certain average along the RG flow. The dilaton is naturally light whenever 
i) confinement is reached fast enough (such as via the condensation of $\OO$) and 
ii) the beta function is small ({\em walking}) at the condensation scale.
These conditions are satisfied for a class of models with a bulk pseudo-Goldstone boson whose potential is nearly flat at small field and exponential at large field values. Thus, the recent observation by Contino, Pomarol and Rattazzi holds in CFTs with a single nearly-marginal operator. 

We also discuss the holographic method to compute the condensate $\vev\OO$, based on solving the first-order nonlinear differential equation that the beta function satisfies.

\end{abstract}

\newpage 

\tableofcontents

\section{Introduction}

The spontaneous breaking of conformal invariance (SBCI) is interesting for two reasons, both connected with naturalness. First, this symmetry breaking pattern may give rise to a Goldstone boson -- the {\em dilaton} -- which should come with a protected mass. Second, the effective theory for the dilaton presents a toy version of the Cosmological Constant (CC) problem that can shed new light on the actual CC problem. Recently, there has been significant progress on these questions due to  Contino, Pomarol and Rattazzi (CPR), who realized that SBCI and a light dilaton should occur naturally in certain nearly marginal deformations of  Conformal Field Theories (CFTs)  \cite{CPR}. CPR suggested a simple holographic implementation that has been checked to work in \cite{Bellazzini:2013fga,Coradeschi:2013gda}. Thus, this represents a first natural model of SBCI and of a light dilaton. This holographic model is dual to a CFT that contains at least two scalar operators. The purpose of the present paper is to examine how the CPR proposal operates in CFTs with a single operator. But first let us review some basic facts.

If a CFT exhibits SBCI (and thus a massless dilaton), then physics at low energies can be captured by EFT methods. The effective theory for such putative light dilaton is constructed by writing down the most general Lagrangian invariant under the scale-transformation $x^\mu \to b^{-1} \, x^\mu$. The canonically normalized dilaton field $\chi$ must be proportional to the (appropriate power of the) scale responsible for SBCI. Therefore, the scale transformation acts like  $\chi\to b^{(d-2)/2} \chi$, in $d>2$ space-time dimensions. 
The effective Lagrangian compatible with this symmetry allows a potential of the form 
$$
V_\dil^{CFT}(\chi)=U_0 \,\chi^{2d\over d-2}~,
$$
with an {\em arbitrary} `quartic' coupling  $U_0$ --  the analogue of the Cosmological Constant. Since any $U_0$ is allowed by the symmetries, a generic CFT is expected to lead (if anything) to a nonzero $U_0$ (for  CFTs with a stable  ground state, $U_0\geq0$). 

In this language, having SBCI means that $\vev\chi\neq0$. But with this potential for $\chi$ it is clear that this only happens in CFTs if the dilaton quartic vanishes, $U_0=0$. Since this value of $U_0$ is not protected in the absence of additional assumptions, this is a fine-tuning. 
Thus, a CFT may exhibit SBCI only if either i) it is fine-tuned  or ii) it enjoys additional symmetries (such as SUSY) that enforce $U_0=0$. Conversely, generic (non-tuned, non-supersymmetric) CFTs do not exhibit SBCI nor a massless dilaton.\footnote{Indeed, 
if $U_0>0$, then $\vev\chi=0$, which means no SBCI and there cannot be a dilaton in the first place, so we reach a contradiction. Thus, for CFTs, $U_0\neq0$ is incompatible with SBCI and with a light or massless dilaton.}

The analogy with the CC problem is also obvious from the above effective potential. 
Anything like a SBCI and a light dilaton must have small $U_0$, but the naive generic EFT expectation is that $U_0$ is large (of order one) at least for strongly coupled theories. Thus, for a light dilaton to be naturally realized something similar to the screening of a large `dilaton constant' $U_0$ must take place.  

The next logical question, raised in \cite{CPR}, is: can there be a naturally light dilaton in QFTs that are {\em close} to conformally invariant (CI)? 
The simplest and closest to a CFT is a CFT deformed by a nearly-marginal operator (`nearly-marginal deformation' for short). This introduces a {\em small} explicit breaking of CI and now the dilaton action does not need to be scale invariant. If the deformation is nearly-marginal, then the effective action should be close to the CI case. To fix terminology, we assume that  a CFT contains a (marginally) relevant operator $\OO$ of dimension ${\rm Dim}(\OO) = {d-\eepsilon}$, and we consider the Lagrangian deformation  defined by adding the perturbation
\beq\label{deformation}
\delta \LL = - \lambda\;\OO~.
\eeq

Since the dilaton shifts like a scale transformation, RG computations capture exactly the dependence on the dilaton. Thus the effective potential is obtained by evaluating the running coupling constants at the scale  $\mu=\chi^{2/(d-2)}$. For the above deformation, this generally takes the form
\beq\label{Vdil}
V_\dil(\chi) = \chi^{2d\over d-2}\;U\big(\lambda(\chi)\big)  \,,
\eeq
where $\lambda(\chi)$ is the running coupling constant $\lambda(\mu)$ (equal to $\lambda_\UV$ at $\Lambda_\UV$). The function $U(\lambda(\chi))$  depends on the model, and we can call it the `dilaton quartic function'.
For small coupling $\lambda$, it should admit a perturbative expansion $U(\lambda)=U_0+U_1 \lambda+\dots$.

Now one can ask whether  SBCI occurs in addition to the explicit breaking of CI: that amounts to having a nontrivial minimum of $V_\dil$. Let us call the scale associated to this vev the {\em infrared} scale, and the location of the minimum $\chi_\IR$.
Since $\chi V_\dil'={2\over d-2}\chi^{2d\over d-2}\left( d\; U + U'\beta\right) $
it follows that if  $\beta\equiv {d\lambda\over d\log\mu}$ at the minimum is small, then both the dilaton quartic at the minimum and the dilaton mass are small. Indeed,
\beq\label{smallVm}
V_\dil|_{\IR}  \equiv U_\IR \chi^{2d\over d-2}_\IR
\sim \beta_\IR\;\chi^{2d\over d-2}_\IR 
\qquad {\rm and}\qquad 
m_\dil^2 = V_\dil''|_\IR \sim \beta_\IR\;\chi_\IR^2
\eeq
(as well as all higher-derivatives of $V_\dil$) are generically suppressed by one power of $\beta_\IR$. 

Note that this conclusion only relies on 2 assumptions: i) the beta function is small (`walking') at $\Lambda_\IR$, the scale where CI is broken; 
ii) the function  $U(\lambda)$ is generic -- it does not contain small or large parameters. 
Of course, it is possible to have additional suppressions by tuning $U(\lambda)$. For instance, if both $U$ and $U'$ vanish at the minimum\footnote{As is arranged for if $U(\lambda)=f(\lambda)^2$ and $f$ has a simple zero \cite{Rattazzi:2000hs,Csaki:2000zn}.}, then $m_\dil^2\sim \beta_\IR^2$. 

Clearly, this has an impact on the two naturalness issues: the dilaton mass and the value of the potential at the minimum can be naturally suppressed at the IR scale, even if it involves order one parameters in the UV. SBCI {\em can} be naturally realized in nearly-marginal deformations. Interestingly enough, in deformed CFTs the  physics of the dilaton is discontinuous in the parameter $\eepsilon$ for $\eepsilon\to0$.

One holographic realization of this idea that was advocated in \cite{CPR} and elaborated in \cite{Bellazzini:2013fga,Coradeschi:2013gda} is also quite simple. The nearly-marginal deformation is realized by a domain wall solution of a nearly-massless scalar $\phi$. The approximate global shift symmetry enjoyed by $\phi$ is responsible for the walking, {\em i.e.}, for protecting a small beta function throughout the flow. 
The dilaton is realized by a hard IR brane that finds its location along the domain wall geometry. 
In field theory language one the hard IR brane corresponds to one (or more) additional  operator(s) $\OO'$ with arbitrarily large dimension $\Delta'$, which develops a condensate -- thereby breaking conformal invariance spontaneously. The dilaton is identified as $\vev{\OO'}^{1\over\Delta'}$, and holographically it corresponds to the (warp factor at the) location of the IR brane, $\chi \equiv a(y_\IR)$. The IR brane is allowed to break the global shift symmetry by the presence of a localized potential on the IR brane $U_{IR\;brane}(\vp)$. In field theory, that is an order-one crossed-coupling between $\OO$ and $\OO'$. 
In the probe limit, the effective potential for the dilaton is the induced metric times the IR potential, $V_\dil = a_\IR^4 U_{IR\;brane}(\vp(a_\IR))$, which realizes exactly  \eqref{Vdil} and leads to identify the IR brane potential $U_{IR\;brane}(\vp)$ as the dilaton quartic function\footnote{One can see that  for nearly-marginal deformation this is true up to small corrections.}. 

Encouraged by this match, in this paper we try to test the CPR proposal in slightly simpler and perhaps more `realistic' models, involving a single nearly marginal operator $\OO$. The RG flows triggered by a relevant deformation already have all the ingredients to possibly realize SBCI, because upon performing a deformation $-\lambda\, \OO$ generally one expects that $\OO$ itself develops a condensate, $\vev\OO$. The fluctuation of $\vev\OO$ then can already play the role of a (possibly light) dilaton. This can be quite easily tested using the techniques developed in the context of holographic RG flows \cite{DeWolfe:1999cp,Bianchi:2001de,de Boer:1999xf,Anselmi:2000fu,Papadimitriou:2004ap,Papadimitriou:2004rz,Papadimitriou:2007sj} and soft-wall models \cite{Gubser:2000nd,Karch:2006pv}. 

Specifically, we consider {\em confining} RG flows, {\em i.e.}, flows that generate a `universal' mass gap $\Lambda_\IR$, in the sense that all operators in the theory acquire gapped spectra controlled by $\Lambda_\IR$. It is rather straightforward to see that this implies that the beta function must go to an order-one constant. 
This seems to jeopardize the possibility that there is a light dilaton -- the beta function cannot be small in the deep IR. However, there is a subtle way out: 
the dilaton will turn out to be light whenever the beta function turns to the confining regime fast enough. The natural and generic way that the beta function experiences a fast rise is by the condensation of $\OO$ itself. The only requirement, then, will be that  the beta function stays small ({\em walking}) at the condensation scale.  The end result will be  completely in line with the CPR observation.

Before going to the main points, a brief note on related literature. 
The interest for the dilaton as the pseud-Goldstone boson of spontaneous breaking of scale has a long history, dating back to \cite{Gildener:1976ih,Yamawaki:1985zg,Bando:1986bg}. More recently there has been a renewed interest mostly triggered by LHC phenomenology -- see \cite{Goldberger:2008zz,Appelquist:2010gy,Grinstein:2011dq,Vecchi:2010gj,Low:2012rj,Chacko:2012sy,Chacko:2012vm,Bellazzini:2012vz,Chacko:2013dra,Serra:2013kga} and references therein. 
The identification of a dilaton in a holographic RG flows goes back at least to \cite{Bianchi:2001de}, which identified a massless dilaton in the tuned SBCI flows and no dilaton in deformation flows. Recently, there was a renewed interest, though some of the cases studied involve what in the present paper we classify as tuned SBCI  \cite{Hoyos:2013gma,Bajc:2013wha}, which displays an exactly massless dilaton. Works discussing the formation of a condensate in holographic CFT deformations include \cite{Papadimitriou:2004ap,Papadimitriou:2004rz,Papadimitriou:2007sj,Kiritsis:2014kua}.
There is also a considerable amount of literature on string-theoretic embeddings of walking dynamics and light dilatons, starting from \cite{Elander:2009pk} (see also \cite{Elander:2013jqa}  and references therein for a more recent account). Since these scenarios are supersymmetric, it is hard to disentangle whether the dilaton lightness stems from SBCI or from SUSY. In the present work, we consider non-supersymmetric flows.

The rest of this paper is organized as follows. In Sec.~\ref{sec:beta}
we review the holographic RG flows, using the language of the
holographic beta function, which is very convenient to understand when
the dilaton should be light. In Sec.~\ref{sec:beqn} we introduce the
differential equation that the holographic beta function obeys, and
describe how it gives a method to compute the condensate $\vev{\OO}$
once the model (the bulk potential) is specified. This method may be
well understood by the expert reader, but we find that a clear and
systematic presentation of it was lacking in the literature. In
Sec.~\ref{sec:flows} we describe the most basic general features of
the flows of main interest, running between a CFT in the UV and a
confining IR and how the condensation of $\OO$ emerges in this
picture.  We then introduce in Sec.~\ref{sec:models} the two
representative models (model `A', which has $\vev\OO\neq0$; and model
`B', with $\vev\OO=0$), for which we will compute the scalar glueball
spectrum.  In Sec.~\ref{sec:spectrum} we present the analytic and
numerical computations of the spectra for both `A' and `B' models. We
offer some conclusions and discussion in Sec.~\ref{sec:concl}.

\section{Nearly-marginal deformations, holographically}
\label{sec:nearly_marginal_def}

In this Section, we review the holographic description of the CFT deformations, and exploit the holographic beta function (or the equivalent superpotential method) to describe how confinement and the condensation of operators arise in holographic RG flows.

Holographically, the deformations of a CFT by a scalar operator $\OO$ are realized as  a domain wall geometry, where a $(d+1)$-dimensional scalar field $\vp$ develops a profile in the `holographic' extra dimension $y$ and the metric responds according to the $(d+1)$-dimensional Einstein equations. To fix the notation and normalizations we shall take the $d+1 $ action as
\beq\label{action}
S= {M^{d-1} \over 2} \int d^{d+1}x \sqrt{-g}\left\{  R - (\partial\phi)^2  - 2 V(\phi)\right\} 
\eeq
where we factor out an overall Planck constant and we will work with dimensionless scalar
$\phi=\phi_{canonical} / M^{(d-1)/2} $ and rescaled potential $V= V_{canonical} / M^{d-1} $. We intend to keep the discussion as independent as possible of the specific choice of the potential $V(\phi)$. However, the main focus will be on monotonous  potentials that admit an AdS vacuum, that we will choose to be at $\phi=0$. 

A `domain wall' geometry is a solution with $d-$dimensional Poincare symmetry, 
\bea\label{background}
ds^2 &=& dy^2 + a^2(y)dx^\mu dx^\nu \eta_{\mu\nu} \,, \cr
\phi&=&\phi(y)~.
\eea
With this type of ansatz, the equations reduce to one 2nd order and one 1st order equations, 
\bea\label{KGbar}
\ddot\phi+d \,{\dot a \over a}\,\dot\phi &=& \barr V'(\phi)\\[2mm]
{d(d-1)\over2}\left({\dot a \over a }\right)^2 &=&{\dot\phi^2\over2} -\barr V(\phi)
\eea
where a dot stands for a $y$-derivative. As it is well known \cite{DeWolfe:1999cp}, these equations can be equivalently written as the following system of first order equations
\beqa\label{1storder}
\dot\phi&=&-\barr W'(\phi) \\[2mm]
{\dot a\over a}&=&{ \barr W(\phi) \over d-1}
\eeqa
in terms of a  {\em superpotential} $W(\phi)$ that obeys
\beq\label{spot2}
\barr W'(\phi)^2=2\barr V(\phi) +{d\over d-1} \barr W^2(\phi) ~.
\eeq

In non-supersymmetric theories, the superpotential equation should be understood as an equation for $W(\phi)$ that depends only on the choice of $V$. Since it is a first order ODE, $W$ is determined up to an integration constant (to be identified as the condensate $\vev{\OO}$ in our case). Below we describe a well defined prescription to fix this integration constant from a certain regularity condition in the IR.

Before that, though, let us bring up two convenient coordinates for the domain wall geometry. The first one is the `Unitary gauge', where the field $\phi$ (being a monotonous function along the geometry) is itself the coordinate. In this coordinate, the metric reads
$$
ds^2 = {d\phi^2 \over [\barr W'(\phi)]^2} + a^2(\phi) dx^\mu dx_\mu \,,
$$
and the only evolution equation is for $a$, $\partial_\phi \log{a} = -{W(\phi) \over (d-1)W'(\phi)}$.

\subsection{The holographic beta function}
\label{sec:beta}

The other convenient choice is the `RG-gauge' where the warp factor $a$ is the coordinate. Since the warp factor $a$ is what plays the role holographically of the renormalization scale $\mu$, in this gauge the comparison with the field theory language is most convenient.

In this gauge, the metric reads 
$$
ds^2 = {da^2 \over a^2 \barr W^2(\phi(a))} + a^2 dx^\mu dx_\mu~.
$$
Once the RG scale $\mu$ is identified with the warp factor, and since the value of $\phi$ close to the boundary is identified as the coupling strength of the deformation (the $\lambda$ in Eq.~\eqref{deformation}),
 one  immediately identifies the holographic version of the beta function \cite{Behrndt:1999ay,de Boer:1999xf,Anselmi:2000fu} as 
\beq\label{beta}
 a \partial_a \phi  =-(d-1){\barr W'(\phi) \over \barr W(\phi)} \equiv \beta(\phi) \,.
\eeq

The results of Section \ref{sec:spectrum} support that this is the correct identification of what plays the role of the beta function. It follows that the superpotential relates to $\beta(\phi)$ as
\beq\label{Wbeta}
\barr W(\phi) ={ d-1\over \ell}\exp\left({-\int_0^\phi d\phi'\; {\beta(\phi') \over d-1}}\right)
\eeq
and that the superpotential equation in terms of the $\beta$ function is the following integral equation~\cite{Galow:2009kw,Veschgini:2010ws} 
\beq\label{spotbeta}
\left(\beta^2  -{d( d-1) } \right) {\barr W^2 \over 2(d-1)^2}=\barr V   \,.
\eeq
In principle one could use this equation (together with \eqref{Wbeta}) to work out $\beta$ once $V$ is given or vice-verse. However, it is slightly more convenient to use \eqref{spot2} first and then \eqref{Wbeta}.

From  Eq.~\eqref{spotbeta} it is clear that the domain walls (with $V(\phi)<0$ everywhere) map to flows with a bounded beta function, 
\beq\label{betabound}
\beta^2 < d(d-1)
\eeq
everywhere along the flow. Also  clear at this point is that a constant beta function maps into exponential superpotential and potential. Shortly, we will see that confining geometries require that for $\phi\to\infty$, $\beta$ goes to an order-one constant $\beta(\phi)\to\beta_\infty$ in the range ${\sqrt{d-1}} < -\beta_\infty < \sqrt{d(d-1)}$.
Hence, we encounter one important difference between the (confining) soft walls and the probe brane realization of \cite{CPR,Bellazzini:2013fga,Coradeschi:2013gda}:  $\beta_\IR$ is not small for the {\em confining} soft walls, and it is a priori unclear from the CPR argument whether one should expect a light dilaton or not. As we will see, the answer is still yes for certain types of flows.

\subsection{The beta function equation}
\label{sec:beqn}
 
Before turning to specifying the flows, let us emphasize an important point. 
Using \eqref{spotbeta} and its derivative, it is straightforward to realize that
the beta function $\beta(\phi)$ which describes the flow obeys a differential equation,
\beq\label{betaEqn}
\beta\,\beta' = (\beta^2-d(d-1))\left({\beta\over d-1}+{1\over2}{V'\over V}\right)~,
\eeq
which we will refer to as {\em the beta-function equation}.\footnote{Equations equivalent to \eqref{betaEqn} have appeared  previously in the literature, {\em e.g.} in \cite{Papadimitriou:2007sj,Bourdier:2013axa}. After the completion of this work we realized that \cite{Dvali:2011uu} also discusses a differential equation for a beta function in a spirit very similar to the following discussion.}

In non-supersymmetric theories, the logic that we will take is that the bulk potential $V(\phi)$ (which is basically a free function within certain limits) is given. And then we solve \eqref{betaEqn} to find $\beta(\phi)$. 
The solutions are typically non-analytic and involve an integration constant, which maps to the condensate $\vev{\OO}$.  The way to fix it -- to compute the condensate-- requires some physical boundary condition, which we are going to take as requiring that the flow is `regular' (minimally singular, rather) in the IR end of the flow. We describe in detail this procedure in the following subsections.

At this stage, it is already clear that \eqref{betaEqn} has a number of notorious features. Fixed points ($\beta=0$) indeed map to extrema of $V$ (Eq.~\eqref{spotbeta} requires them to be AdS extrema). Another type of flows stands out: those with a nonzero but constant $\beta(\phi)$, which give a relaxed notion of self-similarity. Indeed, a constant $\beta(\phi)$ implies that the 5D metric differs from AdS by a {\em scaling} conformal factor ($ds^2 = z^{2\theta} ds_{AdS}^2$ with $\theta\neq0$ and $z$ the conformal coordinate). This is known as ``hyperscaling violation" in the literature \cite{Dong:2012se} (see also \cite{Charmousis:2010zz,Gouteraux:2011ce}),\footnote{The identification of hyperscaling with a constant $\beta$ (of some scalar operator) seems to apply straightforwardly also in  Lorentz non-invariant cases.}. As will become clear in Section~\ref{sec:flows} the hyperscaling `points' with large enough beta function exhibit confinement. 

It is also clear that there are two attractors dominating Eq.~\eqref{betaEqn}, corresponding to the vanishing of the two factors on the r.h.s. of \eqref{betaEqn}. The first one ($\beta=-\sqrt{d(d-1)}$) is an `IR attractor', in the sense that is attractive towards large $\phi$. This is `bad' attractor -- it gives rise to singularities of Gubser's bad type \cite{Gubser:2000nd}. 

The other (`good') attractor depends on the form of $V(\phi)$, and it is a UV attractor (attractive towards small $\phi$). For smooth potentials (specifically, with small $\big({V'\over V}\big)'$), it drives the beta function towards $\beta\simeq - {d-1\over 2} {V'\over V}$, (the  superpotential is driven towards $\propto\sqrt{-V}$). For suitably chosen $V(\phi)$, this then leads to singularities of Gubser's `good' type.  For exponential $V(\phi)$ ({\em i.e.}, $V'/V=$ constant), it leads to constant beta function (hyperscaling flows)\footnote{The scale-covariance of the hyperscaling geometries is directly related to the scale symmetry  enjoyed by the exponential potentials (see e.g. \cite{Garriga:2001ar,Flachi:2003bb}).}. Provided that $V'/V$ is not too large, these are `good' singularities.

Let us finally emphasize that the beta function equation
(\ref{betaEqn}) and the superpotential equation \eqref{spot2} are
basically equivalent problems.  The existence of a smooth, real and
single-valued $W(\phi)$ solution to Eq.~(\ref{spot2}) seems to be
equivalent to existence of an equally smooth solution $\beta(\phi)$ to
the beta equation~(\ref{betaEqn}). This follows from the fact that
given a smooth $\beta$ then $W$ is obtained by
Eq.~(\ref{Wbeta}). Strictly speaking, $\beta(\phi)$ needs to be
integrable, but for most cases of interest this is the
case. Conversely, given a smooth (and differentiable) $W$ then $\beta$
is given by \eqref{beta}.  Thus, for potentials with AdS minima, it is
possible to obtain the usual Breitenlohner-Freedman bound from the
requirement of reality of the solutions to \eqref{betaEqn} (see
Sec.~\ref{sec:flows} below) near the AdS minimum.

For a generic potential $V(\phi)$, it is not possible or easy to show analytically the existence of a smooth solution to the beta function  equation. However, Eq.~\eqref{betaEqn} is straightforward to integrate numerically, for instance by shooting from the `good' attractor (whenever present). 
In the examples that we consider below, $V(\phi)$ interpolates between $const +\phi^2$ for small $\phi$ to an exponential for $\phi\to\infty$ that admits a `good' attractor. For large $\phi$, then, a smooth $\beta(\phi)$ is granted to exist. It is non-trivial, though, that this can be matched to the AdS asymptotics at $\phi=0$, since this requires that $\beta(\phi)$ goes linearly in $\phi$ (equivalently, $W\sim const+ \phi^2$) near $\phi=0$. Typically, if the transition to the exponential behaviour starts at small enough $\phi$, then this becomes incompatible with the AdS asymptotics ($\beta$ is forced to go to a nonzero constant or equivalently $W$ to go like const$+ \phi$) or with single-valuedness. However, it is possible to see numerically that arranging the transition to the exponential regime in $V(\phi)$ to start at large enough $\phi$, then the regular AdS asymptotics can be reached simultaneously with the good singularity attractor at $\phi\to\infty$. 
This is the form of $V(\phi)$ that we consider in all the examples below. For these models, then, one can say that there is numerical evidence that smooth, real and single-valued $\beta(\phi)$ and $W(\phi)$ exist.

We will perform a more detailed study of the implications of \eqref{betaEqn} in Sec.~\ref{sec:flows} for the specific potentials of interest in this work. 

\subsection{CFTs with confining deformations}
\label{sec:flows}

Let us now describe in detail the type of RG flows that most clearly illustrate when and how a naturally light dilaton emerges. To make the discussion as well defined as possible, we will concentrate on flows from close to a CFT in the UV down to a confining IR. 
In this case, the spectrum is granted to be gapped and the dilaton is going to be easily identifiable as a low-lying mass-eigenvalue. 
The simplest realization then requires a CFT with a (marginally) relevant and confining operator $\OO$, which is dual to a scalar field $\phi$ in the bulk. The RG flow of interest will then be given by the deformation of the CFT by the operator $\OO$, and that this deformation is confining maps into having an appropriate form of the bulk potential $V(\phi)$ at large $\phi$.

Some of the main points presented in this section have been derived previously {\em e.g.} in \cite{Papadimitriou:2007sj,Gursoy:2007cb,Gursoy:2007er,Cabrer:2009we,Galow:2009kw,Megias:2010ku,Bourdier:2013axa} and may be  well understood by many readers. Still we find that a concise and hopefully pedagogic  presentation may be useful. 
 
\subsubsection*{The UV}

RG flows from a UV CFT fixed point translate into Domain Wall geometries that are asymptotically AdS (towards the AdS conformal boundary), so we require the potentials $V(\phi)$ to have an AdS extremum. With no loss of generality, we will choose the field variable such that such extremum is at $\phi=0$. Small $\phi$ is then identified as corresponding to `the UV' region. To simplify the analysis, we will restrict to  potentials that are analytic in $\phi$ and even under $\phi\to-\phi$, that is they admit a series expansion in powers of $\phi^2$. For small $\phi$ then we have
\beq\label{V}
V(\phi) = -{d(d-1)\over 2\ell^2} + {M_\phi^2 \over 2} \phi^2 +\OO\big(\phi^4\big)~.
\eeq

Towards the AdS boundary (recall that the $a$ coordinate plays the role of the RG scale $\mu$) such a scalar contains two {\em modes},
\beq\label{modes}
\phi(a) \simeq \lambda \;a^{-\Delta_-} + \vev{\OO} \; a^{-\Delta_+}~. 
\eeq
The sub-leading mode is identified with a CFT operator $\OO$ and the leading one is identified with its conjugated source (or coupling) $\lambda$. This identification is known as {\em standard quantization}, and is the only choice consistent  with unitarity for  nearly-marginal operators. Let us emphasize that the bulk field variable plays the role of a generalized version of the running coupling constant that includes not only the running coupling $\lambda(a) \sim \lambda\,a^{-\Delta_-}$ but also possibly the `running condensate' $\sim \vev{\OO} a^{-\Delta_+}$.

In fact, once the potential is fixed to \eqref{V}, all of this information is nicely encoded in the beta-function equation  \eqref{betaEqn}. One only needs to plug the expansion \eqref{V} in \eqref{betaEqn}, and work out the expansion of $\beta(\phi)$ for $\phi\to0$. Importantly, since $\beta(\phi)$ is only asked to solve \eqref{betaEqn},
one should allow that the beta function contains both a perturbative and a non-perturbative part\footnote{Needless to say, in light of \eqref{beta} or \eqref{Wbeta}, everything that one can say about $\beta$  can immediately be translated in terms of the superpotential: $W(\phi)$, also, admits a similar separation,  $W(\phi)=W_P(\phi)+W_{NP}(\phi)$.  
} 
\cite{Papadimitriou:2007sj,Bourdier:2013axa}, 
$$
\beta(\phi)=\beta_P(\phi)+\beta_{NP}(\phi)~.
$$
The perturbative part, $\beta_P$, is completely fixed by $V$. Writing 
$V(\phi)=\sum_{n=0}v_{2n} \phi^{2n}$   
and $\beta_P(\phi)=\sum_{n=0}\beta_{2n-1} \phi^{2n-1}$ in \eqref{betaEqn} then the $\beta_{2n-1}$'s  are simple algebraic functions of the $v_{2n}$'s.
At the first order
\beq\label{betaP}
\beta_P(\phi)=  - {\Delta} \;\phi + \dots
\eeq 
and, from \eqref{betaEqn}, $\Delta$ is seen to obey $\Delta(\Delta-d) = M_\phi^2/\ell^2$. Thus, $\Delta$ is set to either of  $\Delta_\pm$ with the familiar expressions
\beq\label{deltapm}
\Delta_\pm = {d\over2}\pm\sqrt{\left({d\over2}\right)^2+M^2_\phi\,\ell^2} \,.
\eeq
All the higher order coefficients $\beta_{2n-1}$ with $n>1$ are given by a single-valued expression involving the  $v_{2n}$ and $\beta_{2n-1}$ up to the same order. Thus all the $\beta_{2n-1}$ are completely fixed up to the binary choice  in $\Delta$.\footnote{Strictly speaking, this requires that $\Delta_\pm$ are irrational numbers, an assumption that we will take throughout this paper but which is not essential for our main results.}

Hence for a given potential $V$, there can be two types of beta function (and of superpotential),  with either $\Delta=\Delta_-$ or $\Delta=\Delta_+$ (called $W_-$-type and $W_+$-type in \cite{Papadimitriou:2007sj}). 
The physical meaning of such beta functions is very different. 
From \eqref{beta} or \eqref{1storder}, it follows that a `$\Delta_+$-type' beta function describes a CFT where the operator $\OO$ condenses, that is a CFT with purely spontaneous breaking of conformal invariance.\footnote{Recall that the holographic beta function defined through (\ref{beta}) measures both the running of the deformation coupling as well as the  dimension of the condensate -- it is a measure of the breaking of scale invariance independently of whether the breaking is explicit or spontaneous. In undeformed CFTs with SBCI (the type $W_+$-flows), the holographic $\beta$ function defined by  (\ref{beta}) is not really a beta function of any coupling -- it just reads the (anomalous) dimension of the condensate.} From the discussion in the introduction, one expects that this situation is either fine-tuned or supersymmetric. We will see this more explicitly below. Instead, a  `$\Delta_-$-type' beta function is a genuine deformation of a CFT by the operator $\OO$ ($\delta \LL=-\lambda \OO$) --  an explicit breaking of conformal invariance. \\

Thus,  $\Delta_+$-type beta-functions  represent CFTs with SBCI -- or simply `SBCI flows'. Conversely, $\Delta_-$-type beta-functions are CFT deformations -- or `deformed CFTs'.~ \\

The next step is to realize that, generically, the deformed CFTs also induce the formation of a condensate: an explicit breaking of conformal invariance also induces spontaneous breaking. 
To see this, we turn to the Non-Perturbative part, $\beta_{NP}$. By definition, this is a function that satisfies the nonlinear ODE (for $\beta_{NP}(\phi)$) obtained by plugging $\beta=\beta_P(\phi)+\beta_{NP}(\phi)$ and $V(\phi)$ in \eqref{spot2}. This problem has been studied at length \cite{Papadimitriou:2007sj} (in terms of superpotentials) and the outcome is quite interesting. Generally, $W_{NP}$ is a non-analytic function of $\phi^2$, and its presence is controlled by an integration constant -- usually referred to as $s$. It is possible to parametrize this integration constant so that $s=0$ means $\beta_{NP}=W_{NP}=0$ \cite{Papadimitriou:2007sj}. When $\beta_{NP}$ is a small correction to $\beta_P$, the differential equation can be linearized and the solution explicitly found. Whenever  $\beta(\phi)\ll1 $ (as happens close enough to $\phi=0$) the expression simplifies to 
\beq\label{betaNP}
\beta_{NP}(\phi) \propto s \; \exp\left( - \int_0^\phi d\phi' \;{d+\beta_P'(\phi') \over \beta_P(\phi')} \right) \,.
\eeq
For the obtained UV behaviour of $\beta_P$ \eqref{betaP}, one sees that $\beta_{NP}\sim s \,\phi^{{d\over \Delta_\pm} - 1} $ (the NP superpotential goes like $W_{NP}\sim \phi^{d\over \Delta_\pm}$), which is generically non-analytic in $\phi$. For $\Delta_+$-type flows the only choice compatible with $\beta_{NP}$ being a small correction in the UV is that $s=0$. Thus SBCI flows do not allow a NP part. This  is a first hint that these CFTs have to be tuned.

Instead, deformation beta functions are compatible with $\beta_{NP}$ because $\phi^{{d\over \Delta_-}-1}$ is subleading to  $\phi$. Not only that, for finite $s$ there is always a region of sufficiently small $\phi$ where $W_{NP}$ is well approximated by \eqref{betaNP}.

Now, the physical meaning of the integration constant $s$ is none other than the value of the condensate $\vev{\OO}$. Indeed, upon using \eqref{beta} or \eqref{1storder} and a $\beta(\phi)$ of the form $ \beta = -\Delta_- \phi +\dots -  s \,\phi^{{d\over \Delta_-}-1} +\dots$ (dots denoting higher integer powers), one obtains
\beq\label{phis}
\phi = \lambda \,a^{-\Delta_-}(1+\dots) + s\;{\lambda^{{\Delta_+\over \Delta_-}} \over \Delta_+-\Delta_-}\;\,a^{-\Delta_+} \;(1+\dots)~.
\eeq
Thus, indeed, the subleading mode (dual to the condensate $\vev\OO$) is proportional to $s$. In brief we will see that for generic choices of $V(\phi)$ the condensate is nonzero.

So far, we have seen i) that the perturbative part of the flow is uniquely fixed by the bulk potential $V(\phi)$; ii) that for given $V$ there are at most two types of flows; iii) that the SBCI-flow does not have an integration constant to be adjusted; and iv) that the deformation flows do have one which is physically the condensate.

\subsubsection*{The condensate}
\label{sec:condensate}

Let us now see how to determine the condensate $\vev\OO$ -- or $s$. The main idea is that $s$ should be adjusted so that the IR part of the flow is regular -- or at least as regular as possible. Generally the flows to a confining IR (or any IR which is not a CFT) correspond to singular bulk geometries, with a curvature singularity in the $\phi\to\infty$ region. However, as pointed out by Gubser \cite{Gubser:2000nd} (see also \cite{Wald:1980jn,Horowitz:1995gi} and \cite{Gursoy:2007cb,Gursoy:2007er,Cabrer:2009we,Faulkner:2010fh}) not all singularities are equally `bad'. The natural IR boundary condition that  supplements the beta-function equation \eqref{betaEqn} is that the $\phi\to\infty$ region is as regular as possible.
This selects uniquely a critical value for the integration constant, referred to as $s_c$ in \cite{Papadimitriou:2007sj}, which maps to the physical value of the condensate $\vev{\OO}$ in the {\em regular} ground state of the deformed CFT. 

In light of the properties of the beta-function equation \eqref{betaEqn}, the procedure to find $s_c$ is straightforward: the behaviour of $\beta(\phi)$ for large $\phi$ is dominated by the {\em bad} attractor of \eqref{betaEqn}. If one `over-shoots' from the UV (taking  $s>s_c$), then the solution falls on the bad attractor, $\beta \to -\sqrt{d(d-1)}$. If one under-shoots ($s<s_c$), then one can see that the solutions do not reach out to $\phi\to\infty$ (it becomes singular at a finite $\phi$). The critical solution is, then the first solution that extends for  $\phi\to\infty$; and the last that does not fall into the `bad' attractor. There are many physically meaningful solutions that are examples of such critical solutions, ranging from extremal (hairy) black holes \cite{Gubser:2000nd,Horowitz:1995gi,Faulkner:2010fh} to the AdS Soliton \cite{Witten:1998zw,Horowitz:1998ha}. 

The procedure to find $s_c$ numerically for a general $V(\phi)$ can be made systematic and efficient by shooting instead from the IR -- since then the `bad' attractor is in fact repulsive. 
A class of potentials that is especially simple to treat is given by potentials that grow exponentially for $\phi\to\infty$, {\em i.e.} with  ${V'\over V}=$ constant. For these, the regular $\beta$ function obeys at $\phi\to\infty$
\beq\label{smoothIR}
\beta \to \beta_\infty \equiv -{d-1\over2}{V'\over V}={\rm const}.
\eeq
Whenever  $\beta_\infty > -\sqrt{d(d-1)}$, such a solution is maximally regular --  it is a `good' singularity according to Gubser's criterium.  
In particular, it follows that in order to possibly admit for IR-regular flows, the  potential cannot grow too large -- it is also restricted to 
\beq\label{VsmoothIR}
{V'\over V} < {1\over2}\sqrt{{d\over d-1}}~.
\eeq

The main point that we want to emphasize now is that generically $s_c$ is nonzero. Typically, the only way to have a maximally regular $\beta(\phi)$ in the IR is to allow for a nonzero integration constant. Of course, it is always possible to design a beta function that has no  NP part -- and that is therefore an analytic function of $\phi^2$ that has smooth IR as in \eqref{smoothIR}. However, these inevitably  descend from  fine-tuned potentials -- these are fine-tuned CFTs. And there are two types of such tuned CFTs: CFTs with SBCI for ($\Delta_+$-type flows)  or deformations of CFTs that don't induce a condensate. Both of these situations are automatically generated in supersymmetric cases -- or, whenever one assumes a superpotential which is  an analytic function of $\phi^2$.\footnote{Of course, this is true both for standard and alternative quantization. 
For the alternative quantization, one interprets the leading mode in $\phi$ as $\vev\OO$ and the subleading as $\lambda$. Then, the $\Delta_-$ solutions have a condensate and $\Delta_+$ solutions don't. In addition, generically the $\Delta_-$ solutions have both a condensate and a source term, so these are deformations. The tuned sub-type of $\Delta_-$ solutions with no subleading mode are the tuned CFTs with SBCI. And the $\Delta_+$ solutions are deformations that induce no condensate. }

Instead, {\em for randomly chosen potentials $V(\phi)$} (which are analytic in $\phi^2$ and with smooth enough IR (as in \eqref{VsmoothIR})), then 
\begin{enumerate}
\item {\em generically there is no SBCI flow (no $\Delta_+$-type beta-function or superpotential) with regular IR.} Typically, $\Delta_+$-type flows fall into the bad attractor and so these solutions are discarded as unphysical. {\em Therefore, generic potentials (or CFTs) do not lead to SBCI.} 

\item {\em there is always a deformation flow (a $\Delta_-$-type beta function or superpotential),  generically with a nonzero  condensate $\vev\OO$ ($s_c\neq0$). Therefore, CFT deformations by an operator $\OO$ generically lead to a condensate $\vev\OO$.}
\end{enumerate}
We won't give here a proof of these statements -- they are almost straightforward consequences of the form of the beta function equation, Eq. \eqref{betaEqn}, equipped with the maximally-regular-IR boundary condition.

Instead, we can illustrate that this is what happens looking at the first of the examples that we will work out below. For the Type A model depicted in Fig.~\ref{fig:A},  $V'/V$ behaves like $\sim\Delta_- \phi$ for $\phi<\phi_{conf}$ and jumps to an order-one constant for $\phi>\phi_{conf}$. The numerical integration to extract $\beta(\phi)$ reveals that immediately after the dip in $V'/V$ (going towards small $\phi$), $\beta(\phi)$ behaves like a $\Delta_+-$type solution which is typically displaced from the origin, that is with $\beta(\phi)$ approaching 0 at $\phi\neq0$. In the generic situation depicted in  Fig.~\ref{fig:A} what happens then is that $\beta(\phi)$ is attracted towards $\Delta_-\phi$, so in the end the solution is of $\Delta_--$type. In order for the beta function to be really of $\Delta_+-$type, one should fine tune $\phi_{conf}$ so that $\beta(\phi)$ `lands' exactly at $\phi=0$ while it is still of $\Delta_+-$type. This clearly illustrates both points 1) and 2) above.

The region where the $\beta(\phi)$ looks like a displaced $\Delta_+-$type solution clearly is to be interpreted as the generation of the condensate $\vev\OO$.
Importantly, this also means that one can clearly distinguish two threshold values of $\phi$: the start of the confining region $\phi_{conf}$ and the start of the condensate-dominated region $\phi_{cond}$. It is obvious from the figure that 
$$
\phi_{cond} < \phi_{conf}~.
$$
Integrating \eqref{beta}, this implies that there is also a clear separation of scales: the condensation scale $a_{cond}=a(\phi_{cond})$ is generically bigger than the confinement scale $a_{conf}$. For the type A model of Fig.~\ref{fig:A}, this separation of scales can be estimated as 
$$
{a_{conf}\over a_{cond}}\sim (\Delta_-)^{1/\Delta_+}~.
$$ 
This picture suggests that the scale of the condensate being
parametrically bigger than the confinement scale may be a rather
general rule. Amusingly enough, in QCD the scale of the gluon
condensate is around $400-500\,\textrm{MeV}$~\cite{Shifman:1978bx},
which is about twice of what is usually taken as $\Lambda_{\textrm{QCD}}$.

Another result implied by this prescription to find $\beta(\phi)$ refers to stability. As is well known, the gravity - scalar bulk theories admit `energy bounds' --  for the given asymptotic boundary conditions, one can establish that there is a minimum energy solution (see e.g. \cite{Amsel:2006uf,Faulkner:2010fh,Amsel:2011km} and references therein). This is granted whenever there exists a solution with a superpotential defined for all $\phi$, which is certainly the case for the critical (with $s=s_c$) flows. This implies that the critical flow is the ground state of the deformed theory and that therefore all excitations around the flow should have positive energy. We will check this property below, as the dilaton mode that we will find always has a positive mass-squared. Notice that this applies even in the $s_c<0$ case. Since 
$s_c\sim \vev\OO\sim {\delta S_\eff \over \delta \lambda}$, this might look like a kind of instability. However, these arguments suggest that there  is nothing particularly  unstable about $s_c<0$.

\subsubsection*{The deep IR}

To complete this discussion, let us now briefly characterize some of the possible types of flow in the IR  in terms directly of the (holographic) beta function. The discussion applies at least to theories with a gravity dual. Similar discussions can be found in terms of the superpotential \cite{Gursoy:2007cb,Gursoy:2007er,Cabrer:2009we}. Depending on the form of $\beta(\phi)$ for large $\phi$, the following IR behaviours stand out:
\begin{enumerate}
\item  {\em fixed point}: $\beta(\phi)$ has a simple zero at finite $\phi$. In this case, the flow is from a UV CFT to an IR CFT, and the spectrum is gapless. 

\item {\em `asymptotic' fixed point}: $\beta(\phi)$ vanishes at $\phi\to\infty$.

\item {\em hyperscaling}:  $\beta'(\phi)\to 0$ (with $\beta(\phi)\to\beta_\infty\neq0$) at $\phi\to\infty$, and  $-\beta_\infty < \sqrt{d-1}$. In this case there is no universal mass gap (see below), so this case is qualitatively distinct from confinement and from a fixed point.   

\item {\em confinement}: $\beta(\phi)\to\beta_\infty$, with $  \sqrt{d-1} < -\beta_\infty < \sqrt{d(d-1)}$. In this case, the theory generates a universal mass gap -- the spectrum of all excitations becomes discretized (see below).  
The case $\beta_\infty=-\sqrt{d-1}$ is marginal, and whether it gives rise to confinement and with what sort of discrete spectrum depends on how $\beta(\phi)$ approaches this constant (see e.g. \cite{Cabrer:2009we} for an equivalent discussion in terms of the superpotential).

\end{enumerate}
This list is not intended to be exhaustive. Other more exotic options such as $\beta(\phi)$ approaching asymptotically a periodic (and bounded) function of $\phi$ seem conceivable\footnote{This is certainly compatible with the null energy condition in the bulk since it only requires the potential (in fact, $V'/V$) to be oscillating.}.

In this work, we are mostly interested in confining flows. A simple holographic criterium for confinement ({\em i.e.} universal mass-gap generation) is that the conformal coordinate 
has a finite range. Placing the AdS boundary at $z=0$, this implies that the location of the IR singularity is at a finite conformal coordinate $z=z_s$. 
In terms of the beta function, this leads to the condition
\beq\label{gap}
z_s = \int_0^\infty {d\phi \over a(\phi) W'(\phi)} 
={\ell \over a_0}\int_0^\infty {d\phi \over -\beta(\phi) }\exp{\left(\int_0^\phi d\phi'
\left\{{\beta(\phi')\over d-1}-{1\over\beta(\phi')}\right\}\right)} < \infty \,.
\eeq
Imposing that  $z_s$ is finite, one then  obtains a bound on the beta function
\beq\label{betaConf}
-\beta_\infty > \sqrt{d-1}
\eeq
($\beta_\infty \equiv \lim_{\phi\to\infty} \beta(\phi)$). Thus, confining flows must inescapably have an order-one beta function in the deep IR. Yet, we will see in Sec. \ref{sec:spectrum} that the dilaton can still be suppressed in some cases.

Provided \eqref{gap} is finite, then the mass-splittings in the spectrum of resonances is close to  (see  {\em e.g.}  \cite{Cabrer:2009we}), 
$$
\Lambda_\IR={\pi\over z_s}
$$
and $\Lambda_\IR$ can be thus identified as the mass-gap.
The marginal case $\beta_\infty = -\sqrt{d-1}$ can lead to more QCD-like trajectories for the resonances $m_n^2 \sim n^{\gamma}$ (with $\gamma$ depending on how  $\beta(\phi)$ approaches the line $\beta =-\sqrt{d-1} $), but we won't consider this case here. 

Our analysis will apply straightforwardly to the family of flows where $\beta(\phi)$ asymptotes to 
$$
\beta_\infty \equiv -\nu \sqrt{d(d-1)}
$$
with 
$$
{1\over \sqrt{d}} <\nu <1~.
$$
All of these cases have a finite mass gap and a tower of resonances with an even spacing, $m_n\sim n$. The upper limit $\nu<1$ stems from requiring that the singularities are of Gubser's good type.

\subsection{Two models}
\label{sec:models}

Having presented most of the background material on holographic RG flows, we are now ready to specify the two concrete choices of $V(\phi)$ that allow to illustrate the physics of light dilatons. Our basic requirements are that 1) the UV corresponds to a nearly-marginal (`walking') deformation of a CFT and 2) that the IR is confining. Thus, the potential $V(\phi)$ needs to have 1) an approximate shift symmetry in the UV region ($\phi\ll1$); 
and 2) $V'/V\sim$ of order one and therefore a large breaking of the shift symmetry in the IR (for $\phi\gg1$). 

At this point let us only note that this kind of potential is natural in the technical sense. The flatness of $V$ is protected by the shift symmetry which is certainly realized for small $\phi$,  and the order-one breaking of the shift symmetry can stay localized to the large $\phi$ region. In effective field theory reasoning, the shift-symmetry breaking terms are irrelevant operators --  the shift symmetry in the bulk can be seen as an `emergent' or accidental symmetry.

The two models below will share these properties. Their difference will be in the presence/absence of the condensate $\vev\OO$.

\subsubsection*{Model A: $\vev\OO\neq0$}
\label{sec:modelA}

\begin{figure}[t]
\begin{center}
  \includegraphics[width=12cm]{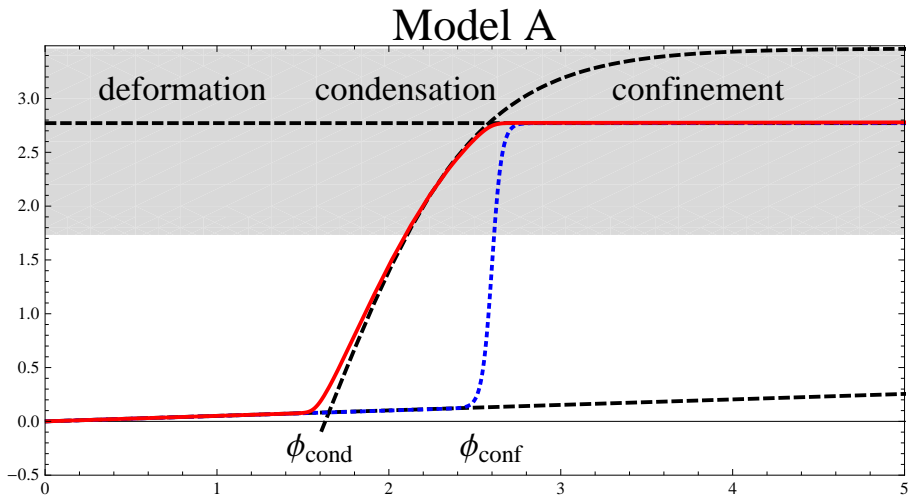}
\end{center}
\caption{Form of ${d-1\over2}\,{V'\over V}$ (blue dotted) and 
$-\beta(\phi)$ (continuous red) for the model `A'. Three regimes (deformation, condensation and confinement) are clearly distinguishable in $\beta(\phi)$.  The black dashed lines are the behaviour of the beta function in the three regions. In the deformation region, $-\beta(\phi)\simeq \Delta_- \phi$. In the condensate-dominated region it is given by the $\Delta_+-$type form $\beta\simeq -\Delta_+ (\phi-\phi_{cond})$, or even better, Eq.~\eqref{tanh}; in the confinement region $-\beta(\phi)\simeq \nu \sqrt{d(d-1)} $. 
$-\beta(\phi)$ tracks the function ${(d-1)\over2}\,{V'\over V}$ everywhere except on the condensate-dominated region. The shaded area corresponds to regular and confining flows. The plot is for $\Delta_-=0.05$, $\nu=0.8$, $\phi_{conf}=2.6$, $d=4$ and a smoothing of the jump in $V'/V$ with $N=20$.}
\label{fig:A}
\end{figure}

Let us start with a representative model of the most generic case, when the condensate $\vev\OO$ is present. The only thing that we need to do is to take the simplest potential $V(\phi)$ that joins an approximately shift-invariant UV with a confining IR. In terms of \beq\label{Vv}
V(\phi)\equiv -{d(d-1)\over \ell^2}\, {\rm exp}\left({v(\phi)}\right)
\eeq
we need $v'(\phi)$ small for $\phi\ll1$ and $v'(\phi)>{1\over2}{d\over d-1}$ for $\phi\gg1$. 
Thus, we will consider $v'(\phi)$ to experience a sharp order one jump at a certain $\phi_{conf}$,
\beq\label{vJump}
v'(\phi) =\begin{cases} 
{2(M_\phi\ell)^2 \phi \over d(d-1) } & {\rm  for} \quad \phi\ll\phi_{conf} \\
{2\nu \over \sqrt{d-1}}& {\rm  for}\quad \phi \gg \phi_{conf} 
\end{cases}
\quad\Rightarrow\quad
v(\phi) =\begin{cases} 
{(M_\phi\ell)^2 \phi^2 \over d(d-1) } & {\rm  for} \quad \phi\ll\phi_{conf} \\
{2\nu \over \sqrt{d-1}}\phi+v_0& {\rm  for}\quad \phi \gg \phi_{conf} 
\end{cases}
\eeq
The precise form of $V(\phi)$ is not important so long as it satisfies these properties.  An analytic expression for this type of potential can easily be obtained, {\em e.g.}, like
$V(\phi) = -{d(d-1)\over \ell^2} \left(V_{\UV}^N + V_\IR^N \right)^{1\over N}$, with $V_{\UV/\IR}$  having the corresponding UV/IR behaviours with \\ $V_{\UV}(\phi_{conf})= V_{\IR}(\phi_{conf})$, and with a large enough $N$.

One then has to integrate \eqref{betaEqn} to extract $\beta(\phi)$. Shooting from the IR (large $\phi$) close to the good attractor $\beta(\phi)\simeq - \nu \sqrt{d(d-1)}$ immediately leads to the smooth beta function displayed in Fig. \ref{fig:A}. 
We see that $\beta(\phi)$ stays close to the attractor $- \nu \sqrt{d(d-1)}$ until $v'$ experiences the jump. The beta function then enters a `condensate-dominated' region. In this region, $\beta(\phi)$ is well approximated by a  condensate-type ($\Delta_+$-type) solution which is displaced from the origin. It goes like $\beta\simeq -\Delta_+ (\phi-\phi_{cond})$ with some constant $\phi_{cond}$ for $\phi-\phi_{cond}$ small. 
A better approximation can be found by solving \eqref{betaEqn} with $V'=0$. At zero-th order in $\Delta_-$, one finds  
\beq\label{tanh}
-\beta(\phi)\simeq\sqrt{d(d-1)}\tanh\left[{d \over d-1}\Big(\phi-\phi_{cond}\Big)\right]~.
\eeq
Notice that this expression re-sums the perturbative expansion in powers of $s$ (the  first term of which is given by \eqref{betaNP}) in the region where the condensate dominates.

Eventually, $\beta(\phi)$ `lands' again on the good attractor and tracks its UV behaviour $\beta(\phi)\simeq -\Delta_- \phi$ until $\phi=0$. An important remnant of the condensate-dominated region is the non-zero integration constant $s_c$ carried by the Non-Perturbative part -- the condensate $\vev\OO$ -- which is generated in this way, the subleading mode $\sim a^{-\Delta_+}$ in \eqref{phis}.

Incidentally, the fact that the $\Delta_+$-type (or SBCI) flows  do not exist generically for a given $V(\phi)$ is clearly seen in this exercise: in order to have such a SBCI flow we would need that $\phi_{cond}=0$, so that the condensate region lands at $\phi=0$. This is clearly non-generic.  

For later reference, let us bring in some convenient labels for the most notorious landmarks of this flow. The value of $\phi$ where the condensate starts to dominate (coming from UV) is what we call $\phi_{cond}$. The value of $\phi$ where we enter the confinement region, we label $\phi_{conf}$ and it is where $v'(\phi)$ experiences the jump. Hence, we see that the condensate dominates before confinement. This is expected  to be generic, and suggests that the scale of the condensates is slightly bigger than that of confinement. 

As we will see, the most interesting case consists of flows that stay in a walking  regime as close as possible to the confining `deep IR' region. This is accomplished by taking $M_\phi \ell\ll1$ and having the jump in $v'(\phi)$ at a moderate $\phi$ ($\phi_{conf}=$ a few).

\subsubsection*{Model B: $\vev\OO=0$}
\label{sec:modelB}

\begin{figure}[t]
\begin{center}
  \includegraphics[width=12cm]{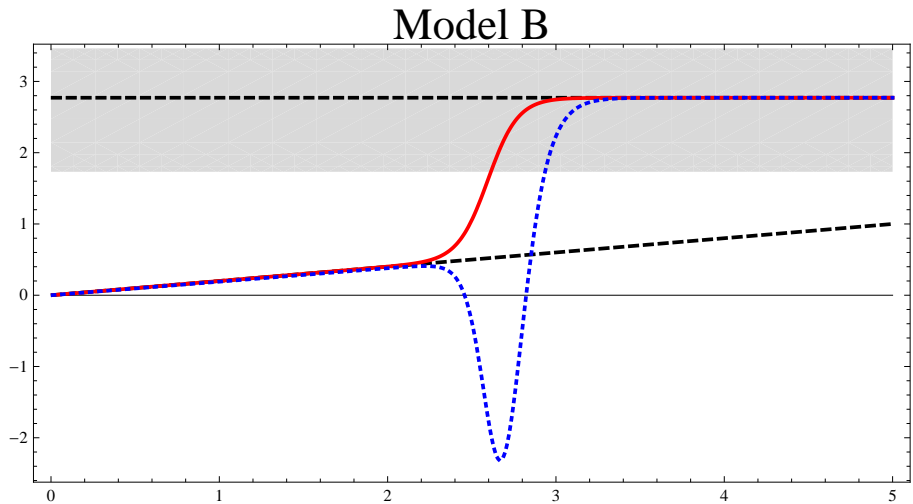}
\end{center}
\caption{Plot of ${d-1\over2}\,{V'\over V}$ (blue dotted line) and of
$-\beta(\phi)$ (continuous red line) for the model `B', with $\Delta_-=0.2$, $\nu=0.8$, $\phi_{conf}=2.6$, $d=4$ and a smoothing of the jump in $\beta$ with $N=15$. The blue dotted line is ${(d-1)\over2}\,{V'\over V}$. The black dashed lines are the behaviour of the beta function in the deformation region and confinement regions. There is still something like a  `condensate-dominated' region where $\beta$ does not track $V'/V$, but the condensate ($\vev\OO\propto s_c$) in this flow vanishes.}
\label{fig:B}
\end{figure}

It is, of course, also possible to design the beta function so that there is no condensate. This is in fact quite easy since this is the case that corresponds to an analytic expression for $\beta(\phi)$. Hence, one only needs to pick an analytic function with appropriate UV and IR behaviour and it is granted that this will be a physical flow with $s_c=\vev\OO=0$ in the ground state.
Needless to say, this is far from being the generic case. 
So, just for the sake of comparison, let us see what happens in this case --  it turns out that the dilaton also notices a tuning.

It suffices to take (as we did previously for $v'$) any smoothed version of a $\beta(\phi)$  piece-wise defined
\beq\label{betaJump}
\beta(\phi) =\begin{cases} 
{-\Delta_- \phi } & {\rm  for} \quad \phi\ll\phi_{conf} \\
{-\nu \sqrt{d(d-1)}}& {\rm  for}\quad \phi \gg \phi_{conf} 
\end{cases}
\eeq
A simple analytic expression is obtained directly in terms of the superpotential, like
$W(\phi) = {d-1\over \ell} \left(W_{\UV}^N + W_\IR^N \right)^{1/N}$, with $W_{\UV/\IR}$  chosen appropriately and with $N$ large enough.

We display in Fig.~\ref{fig:B} the resulting $\beta(\phi)$ and $v'(\phi)$. The oscillating feature in  $v'(\phi)$ near the jump in $\beta(\phi)$ is a manifestation that the potential needs to be carefully adjusted in order to avoid forming a condensate.

\section{Spectra}
\label{sec:spectrum}

In this Section, we present the computation of the scalar spectrum in general for nearly-marginal deformations and specifically for the two representative models introduced in Sec.~\ref{sec:models}. There are in the literature many computations very close to the one in this section, {\em e.g.} \cite{Arutyunov:2000rq,Tanaka:2000er,Bianchi:2001de,Maldacena:2002vr}, though these were not considering precisely the present problem. We will mostly follow Ref.~\cite{Maldacena:2002vr} adapted to the holographic setup.

The most general scalar perturbation of the background flow solution is given by allowing
$$
\phi=\phi_0(y)+\delta\phi
\qquad {\rm and} \qquad ds^2 = N^2 dy^2 + g_{\mu\nu}(dx^\mu+N^\mu dy)(dx^\nu+N^\nu dy)
$$
with $N=1+\delta N$, $N_\mu=\partial_\mu \psi$, $g_{\mu\nu}=a^2(y)\,e^{2\xi}\,\eta_{\mu\nu}$.
In the following, we will work in the {\em Unitary gauge}, $\delta\phi=0$, though everything can be equivalently done using gauge-invariant variables. In this gauge, the dynamically propagating bulk scalar is $\xi$. Since $\xi+\delta\phi / \beta$ is a gauge invariant variable then one can translate from the $\xi$ variable into a $\delta\phi$ perturbation by adding a factor $\beta$. Here and below, $\beta$ stands for the background beta function, $\beta(\phi_0(y))={\dot\phi_0\over (\log a)\dot{}}$~, and we recall that dots stand for $y$-derivatives.

It is straightforward to obtain linearized equations that $\delta N$ and $N_\mu$ satisfy, and one finds that both variables are constrained to be given in terms of $\xi$ and its derivatives. Substituting these expressions into the action  \eqref{action}, using the background equations of motion and integrating by parts a number of times, one quite easily arrives at the following `reduced' action for $\xi$
\beq\label{redaction}
S_{red} = {M^{d-1}\over 2} \int d^{d}x dy \;a^d {\beta^2}\; \left[{(\partial_\mu \xi)^2\over a^2} - \dot\xi^2 \right]  ~.
\eeq
Thus, the equation of motion for $\xi$ in terms of the background proper coordinate $y$ reads
\beq\label{xieom}
{1\over a^d\,\beta^2}\;{d\over dy} \left[a^d\,\beta^2\;\dot\xi\right] - {\Box_0\xi \over a^2} =0~,
\eeq
with $\Box_0=\partial_t^2-\partial_x^2-\dots$.
Introducing a Fourier decomposition in the boundary coordinates $\xi=\xi_n e^{-ik_\mu x^\mu}$, one then obtains an eigenvalue problem for the mode masses $k_\mu k^\mu = m_n^2$,
\beq\label{xieigen}
{1\over a^d\,\beta^2}\;{d\over dy} \left[a^d\,\beta^2\;\dot\xi_n\right] + m_n^2{\xi_n \over a^2} =0~.
\eeq
The spectrum of $m_n$ becomes discrete once we provide boundary conditions at the two ends of the flow. On the UV end (at conformal coordinate $z=0$), we will require that $\delta\phi$ contains only the sub-leading mode (standard quantization), thus $\beta \,\xi \propto z^{\Delta_+}$.
On the IR end (at  conformal coordinate $z=z_s$), we will require that the modes are regular. As will become clear below, one can require that the modes go to a constant at  the singularity, $z= z_s$.

\subsection{A massless dilaton in CFTs with SBCI}
\label{sec:massless}

Before going to the two representative models with a presumably light dilaton, let us first review the case with an exactly massless dilaton, which should be present in the SBCI flows (the $\Delta_+$-type flows), independently of the fact that these are fine tuned. A similar computation can be found in \cite{Bianchi:2001de}.

In fact, it is illustrative to start requiring  that the dilaton is massless and see what this demands on the background flow.
Thus, let us assume that there is one zero eigenvalue. Setting $m_n=0$ in \eqref{xieigen}, one immediately finds (from now on we switch to the conformal coordinate $dz=dy/a$)
\beq\label{masslessmodes}
\xi 
=c_1 + c_2 \int_{0}^z dz' {1\over a^{d-1}\,\beta^2}~.
\eeq
We can now simply check when this expression satisfies all the boundary conditions (and is normalizable). 

Let us start by the IR. Note that if $m_n$ is indeed vanishing, then there are no approximations in \eqref{masslessmodes} -- this expression holds even at the singularity.
Since the $c_2$-mode clearly  diverges at the singularity, then IR-regularity requires $c_2=0$. Hence, when present, a massless dilaton has a constant wavefunction in the $\xi$ variable. 
   
Let us now see when this also satisfies the UV boundary condition. A constant-$\xi$ mode translates into a $\delta\phi$ mode proportional to the background beta function
\beq\label{DWdispl}
\delta\phi  = c_1 \; \beta~.
\eeq
Thus, whether this mode belongs to the spectrum of physical modes depends only on the type of flow. If the flow is of SBCI-type (a `$\Delta_+$-type' flow, with only the sub-leading mode $\phi\sim z^{\Delta_+}$  turned on), then such a $\delta\phi$ mode automatically satisfies `standard quantization' prescription. Since the $\delta\phi$ wave-function involves the fast-decaying mode, it is granted that this mode will be normalizable. Hence the dilaton is a physical massless mode for the SBCI flows.

Instead, if the flow is of the deformation-type (a `$\Delta_-$-type' flow, with the leading mode $\phi\sim z^{\Delta_-}$ turned on), then such a $\delta\phi$ mode does not satisfy `standard quantization' prescription\footnote{For $\Delta_-<(d-2)/2$, in addition, the wave-function of such a mode would not be normalizable.}. This proves that the dilaton cannot be massless in CFT deformations. 

Let us only add at this point that the reader familiar with Domain Walls will find \eqref{DWdispl} quite reasonable. Since $\beta \propto \dot\phi_0$  this is just the mode that displaces the Domain Wall location $\phi(y,x) = \phi_0\big(y+\delta y(x)\big)$ along the {\em holographic} direction. Since the SBCI flows only contain the condensate ($\phi \sim \vev\OO\;z^{\Delta_+}+\dots$), a displacement of the domain wall location then is equivalent to a rescaling of  $\vev\OO$, and this should indeed be massless. 
From this point of view, it is less intuitive why there is no massless dilaton for Domain Walls dual to CFT deformations. From the previous paragraph, the massless Domain Wall displacement mode ($\phi(y,x) = \phi_0\big(y+\delta y(x)\big)$) is still there, and regular in the IR. The problem is that it fails to fulfill (asymptotically) Conformally Invariant boundary conditions and/or normalizability.

\subsection{A light dilaton in marginal CFT deformations}
\label{sec:light}

\begin{figure}[t]
\begin{center}
  \includegraphics[width=10cm]{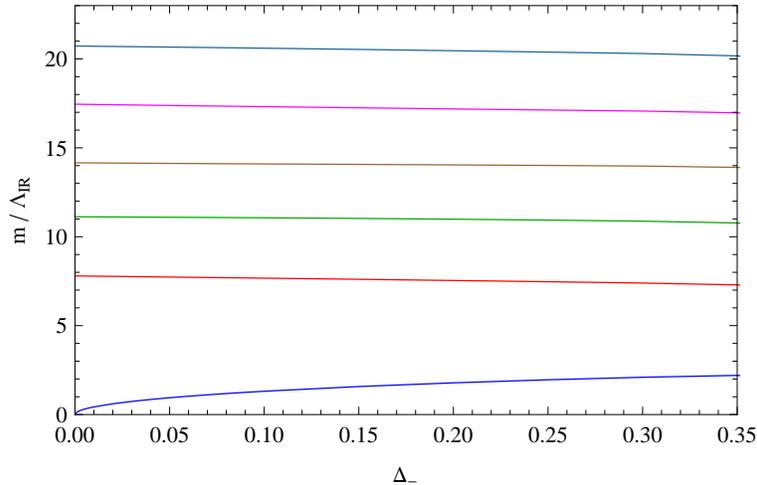}
\end{center}
\caption{Spectrum for the Type-A model as a function of $\Delta_-$ -- the scaling dimension of the nearly-marginal coupling $\lambda$. The first 6 modes are depicted. The lightest mode -- the dilaton -- scales like $m_\dil^2 \sim \Delta_-$.}
\label{fig:spectrumA}
\end{figure}

Let us now see that the dilaton is massive in CFT deformations, and possibly light if the deformations are nearly-marginal.

\begin{figure}[t]
\begin{center}
  \includegraphics[width=10cm]{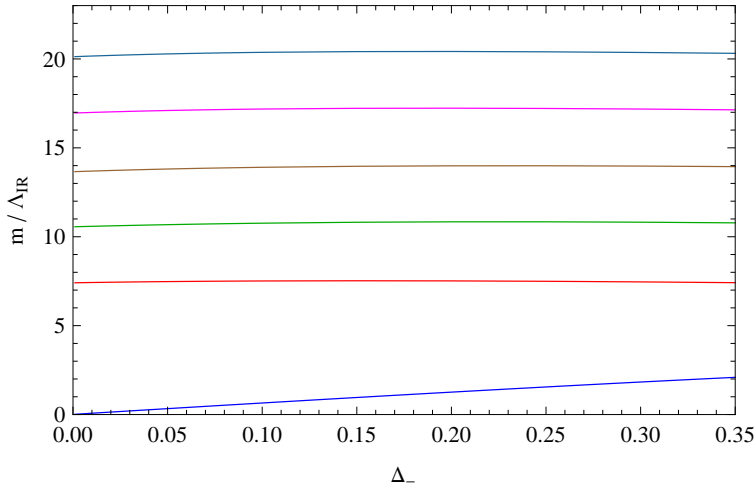}
\end{center}
\caption{Spectrum for the Type-B model as a function of $\Delta_-$. The first 6 modes are depicted. The lightest mode -- the dilaton -- scales like $m_\dil \sim \Delta_-$.}
\label{fig:spectrumB}
\end{figure}

Lacking generically an analytic solution to the eigenvalue problem \eqref{xieigen}, finding the full spectrum requires numerical methods. 
The first few modes for the spectra of the above models `A' and `B' are depicted in Figs.~\ref{fig:spectrumA}  and \ref{fig:spectrumB} respectively. It is obvious from the figures that there is a light state -- the  dilaton is light.

In the following, we concentrate on the light mode, which allows for an analytic treatment\footnote{Whenever there is a mass gap, the $m_n\neq0$ modes separate into `heavy' ($m\gtrsim \Lambda_\IR$) and `light' ($m \ll \Lambda_\IR$) modes.}. The reader not interested in technical details may jump to the mass formula for the dilaton, Eq. \eqref{integral}.

The technique to solve for the light mode is the standard method of matched asymptotic expansions (see {\em e.g.} \cite{Hoyos:2013gma,Bednik:2013nxa}). That is, one finds approximations to the wave-function $\xi_\dil$ separately valid in the UV ($\xi_\dil^\UV$) and the IR regions ($\xi_\dil^\IR$), and matches them together. Implementing the UV (IR) boundary condition on  $\xi_\dil^\UV$ ($\xi_\dil^\IR$) is straightforward, and then the matching procedure determines the eigenvalue, $m_\dil^2$. Importantly, if the mode is light enough ($m_\dil\ll\Lambda_\IR$), then both the UV- and the IR- expansions have a finite overlap region of validity, which allows a well defined matching procedure.

\subsubsection*{UV-expansion}
Treating $m^2$ as a small parameter in \eqref{xieigen}, the solutions in the near-boundary region of \eqref{xieigen} can be obtained iteratively in powers of $m^2$ starting from \eqref{masslessmodes}. Formally, the UV expansion takes the form (see {\em e.g.} \cite{Hoyos:2013gma})
\beq\label{UVexp}
\xi_\dil^{\UV} = \left(1- m^2 \,{\cal I}_\UV +\dots\right)\left[c_1+c_2 \int_{0}^z dz' {1\over a^{d-1}\,\beta^2}\right] \,,
\eeq
with the integral operator (acting to the right) 
$$
{\cal I}_\UV\equiv \int_0^{z} dz' \; {1\over a^{d-1} \beta^2} \int_0^{z'} dz''{a^{d-1} \beta^2}  \,,
$$ 
and the dots denote higher powers of the operator $m^2{\cal I}_\UV$. In the following it will suffice to work at the lowest order in $m$ so one can drop the round brackets in \eqref{UVexp}. But it is important to display the higher order terms so that we can estimate the validity of this approximation.
Note that the expansion parameter really is $m^2/a^2$. Then, the validity of the UV small $m$ expansion is estimated as that $z_m$ obeying $z_s m/a(z_m) \sim 1$.

The UV-boundary condition is `standard quantization', {\em i.e.}, no leading ($z^{\Delta_-}$) mode in  $\delta \phi = \beta\;\xi$. Since we are now assuming that the flow is a CFT-defomation, $\beta \sim \Delta_- z^{\Delta_-}$, we require $c_1=0$,
\beq\label{UVapprox}
\xi_\dil^{\UV} \simeq c_2 \int_{0}^z dz' {1\over a^{d-1}\,\beta^2} \,.
\eeq
Indeed, with $c_1=0$ then $\delta\phi \sim c_2 \;z^{d-\Delta_-}$. We see that $c_2$ represents a shift of the background value of the condensate $\vev\OO$, which looks like a reasonable candidate to play the role of the dilaton.

\subsubsection*{IR-expansion}

The IR-expansion works exactly like the UV-expansion, only that one starts the integrations at the IR end of the flow, that is on the singularity $z=z_s$. Thus we write
\beq\label{IRexp1}
\xi_\dil^{\IR} = \left(1- m^2 \,{\cal I}_\IR +\dots\right)\left[b_1+b_2 \int_{z}^{z_s} dz' {1\over a^{d-1}\,\beta^2}\right] \,,
\eeq
with the integral operator now being
$$
{\cal I}_\IR\equiv \int_z^{z_s} dz' \; {1\over a^{d-1} \beta^2} \int_{z'}^{z_s} dz''{a^{d-1} \beta^2} \,.
$$ 
At zero-th order in $m^2$ the two linearly independent solutions are still of the form \eqref{masslessmodes}, and the regularity on the IR selects $b_2=0$. That is because for the confining models that we consider ($z_s$ finite) the  integral  $\int_{z}^{z_s} dz' {1\over a^{d-1}\,\beta^2}$ is always divergent\footnote{In conformal coordinate, close to $z=z_s$ one has $a(z)\simeq a_s\big({z_s-z\over z_s}\big)^{1\over d\nu^2-1}$ and $\beta\to const$. Thus, the $b_2$ integral is finite only  for $\nu<1/\sqrt{d}$, but in this case $z_s$ is not finite -- there is no confinement. In RG flows with a fixed point or hyperscaling in the IR, instead, one may allow for a nonzero $b_2$ consistently with IR-regularty, suggesting that the IR boundary condition is qualitatively different in those cases. Indeed, for IR CFTs, the relevant condition is the ingoing boundary condition \cite{Dubovsky:2000am}. The dilaton mode should then be realized as a quasi-normal mode -- it should have a finite decay width. We leave this case for future investigation.}.
Thus we now have, 
\beq\label{IRexp}
\xi_\dil^{\IR} = b_1 \left(1- m^2 \,{\cal I}_\IR \right)
\eeq
with ${\cal I}_\IR$ acting on $1$, and we keep the first term in parethesis in order to capture the leading order   $z-$dependence.

Notice that deep enough in the confining region, with $\beta(\phi)=\beta_\infty=const$, and  $W\sim W_0 \, e^{\nu\;\sqrt{d\over d-1}\;\phi}$ one can solve analytically \eqref{xieom} in terms of Bessel functions, of which one selects the regular one. This can be useful for the heavy modes but not for the light dilaton mode, because this approximation inevitably  fails when $\beta(\phi)$ departs from $\beta_\infty$. The approximation \eqref{IRexp}, instead, holds for values of $z$ (and $\phi$) much closer to the UV, and so maximizes the overlap region.

\subsubsection*{Matching}

\begin{figure}[t]
  \begin{tabular}{cc}
      \includegraphics[width=8cm]{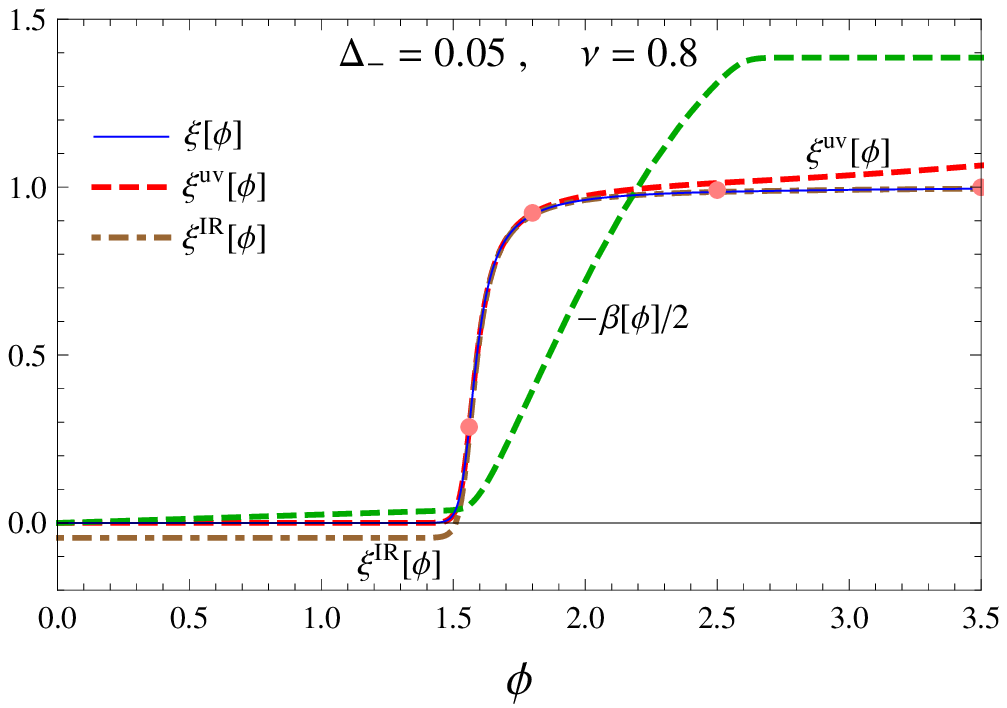}  &
     \includegraphics[width=8cm]{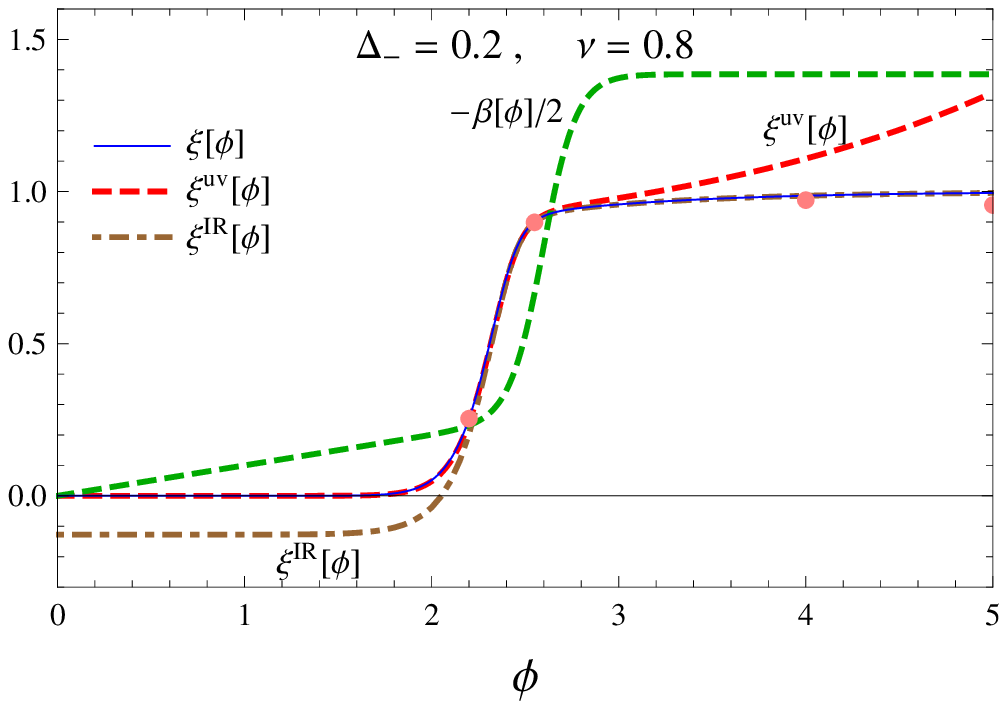}\\
(A) 
          &      
          (B)
  \end{tabular}
\caption{The dilaton wave function $\xi(\phi)$ as a function of the $\phi-$coordinate (continuous blue line), obtained by numerically solving  Eq. \eqref{xieigen}.  The UV- and  IR- approximations to $\xi(\phi)$ (Eqs.~(\ref{UVapprox}) and (\ref{IRexp}) respectively), and the background beta function are also plotted. The dilaton wavefunction switches on at the condensation threshold $\phi_{cond}$.
The pink bullets indicate $\xi_\dil^{\UV}$ computed up to the $m^2 \,{\cal I}_\UV$ term in Eq.~(\ref{UVexp}) at some sample points. This already approximates quite well to the numerical  $\xi(\phi)$, indicating a rather fast convergence of the perturbative expansion. The left (right) panel refers to model A (B). }
\label{fig:wavefns}
\end{figure}

The overlap region where both \eqref{UVexp} and \eqref{IRexp} are valid is approximately given by  ${z_s\over2} \lesssim z \ll z_m $, which is certainly a finite region of the flow for  $z_s m\ll1$, that is precisely when there the dilaton is light. It is now enough to require 
\beq\label{matching}
\xi_\dil^{\UV}(z_\ma)=\xi_\dil^{\IR}(z_\ma)
\qquad {\rm and} \qquad
{\xi_\dil^{\UV}}'(z_\ma)={\xi_\dil^{\IR}}'(z_\ma)
\eeq
at a point $z_\ma$ in the overlap region. The precise choice of $z_\ma$ should not change much the results -- by definition of overlap region -- but is possible to choose $z_\ma$ to improve the approximation done by truncating $\xi^\UV$ and $\xi^\IR$ at a given order (see below). For the moment, one can clearly see in the plots of Fig.~\ref{fig:wavefns} that there is indeed a  considerably wide overlap region (which widens more the smaller is $\Delta_-$).

\begin{figure}[t]
\begin{center}
  \includegraphics[width=10cm]{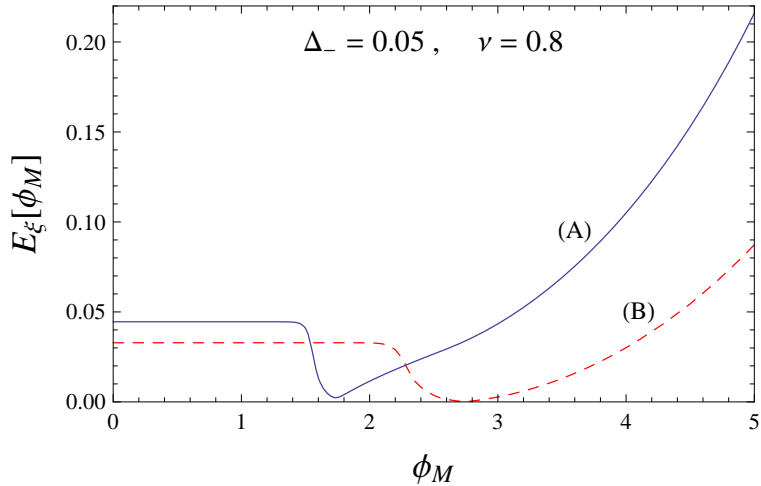}
\end{center}
\caption{Combined residuals in the approximations to the wave function for model A (continuous line) and B (dashed line), defined as $E^2_\xi=(\xi^\UV-\xi)^2+(\xi^\IR-\xi)^2$. The location of the minimum is the optimal $\phi_\ma$. At the minimum  $E_\xi$ is of order $10^{-3}$ and  $10^{-4}$ for the A and B models respectively. The fact that the errors are small in a considerably large region (the wave function $\xi(\phi)$ is normalized to have amplitude 1 at $\phi\to\infty$), indicates that the overlap regions are quite large. The error saturates to a small value for $\phi_M\to0$, which justifies why the $z_\ma\to0$ limit taken in \eqref{integral} is an acceptable approximation.}
\label{fig:error}
\end{figure}

Taking $c_1=0$ and $b_2=0$, the matching conditions \eqref{matching} are easily seen to give
$$
{c_2 \over b_1}\simeq  m_\dil^2 \, \int_{z_{\rm M}}^{z_s} dz \, a^{d-1}\,\beta^2 \,, 
$$
and 
\beq\label{integral1}
{1\over m_\dil^2}\simeq 
\left(\int_{0}^{z_{\rm M}} dz \,{1\over a^{d-1}\,\beta^2}\right)
\left( \int_{z_{\rm M}}^{z_s} dz\,{a^{d-1} \beta^2}\right)
+\int_{z_{\rm M}}^{z_s} dz \; {1\over a^{d-1} \beta^2} \int_{z}^{z_s} dz'{a^{d-1} \beta^2} ~.
\eeq
Interestingly, $m_\dil$ is expressible in terms of integrals over the flow -- some sort of averages of the beta function. Equation \eqref{integral1} still depends on the matching point $z_\ma$, which is to be chosen so that the approximation works as well as possible. This can be decided by plotting the combined error -- the error committed by both the UV- and the IR- expansions added in quadrature. We display the result of the error in Fig.~\ref{fig:error}.

\begin{figure}[t]
\begin{center}
  \includegraphics[width=10cm]{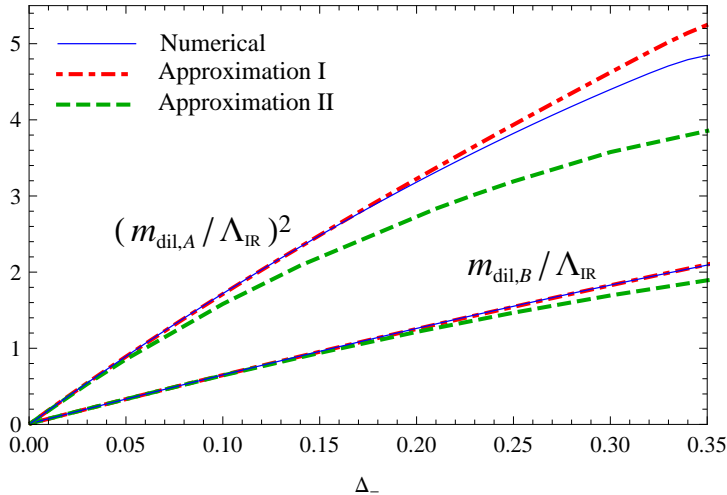}
\end{center}
\caption{Comparison of the numerical values for the dilaton mass (continuous lines) with the two analytic formulae: `Approximation I' refers to Eq.~\eqref{integral1} (dashed-dotted lines) and 'Approximation II' to Eq.~\eqref{integral}. All lines go like $m_{\dil\,A}^2 \sim \Delta_- \Lambda_\IR^2$ and $m_{\dil\,B}^2 \sim \Delta_-^2 \Lambda_\IR^2$.}
\label{fig:formula}
\end{figure}

There is, however, a much more interesting choice to take: $z_\ma =0 $. This does not give as good an approximation but it has the virtue that it provides a formula for $m_\dil$ that does not involve $z_\ma$ anymore, 
\beq\label{integral}
m_\dil^2\simeq{1\over  
\int_{0}^{z_s} dz \; {1\over a^{d-1} \beta^2} \int_{z}^{z_s} dz'{a^{d-1} \beta^2}}~.
\eeq
Remarkably, sending $z_\ma \to 0$ is not only finite, but it is even a reasonably good approximation (see Figs.~\ref{fig:error} and \ref{fig:formula}).

Now, let us use \eqref{integral} to extract when the dilaton is light. The `average' in \eqref{integral} is of a very special kind. 
The structure $[\int a^{1-d}\beta^{-2}]^{-1}$ is automatically small if $\beta$ is small (walking) in  the IR ($a$ small). 

To see more precisely how the double integral behaves, let us consider a toy model that captures the main features of the flow. We can simplify the whole flow by assuming that~$\beta(\phi)$ is a piecewise linear function in 3 distinct regions: deformation,  condensation and confinement. Let us model the  condensate-dominated by a generic slope $\tilde\Delta_+$, which maps to the dimension of the condensate. In this way we allow that the operator dimension in the IR may not be the same as in the UV. 
Thus we take
\beq\label{betatoy}
-\beta(\phi) =\begin{cases} 
\Delta_- \phi  & {\rm  for} \quad 0<\phi < \phi_{cond} \\
\widetilde\Delta_+ (\phi-\phi_{cond})+\Delta_-\phi_{cond} \quad & {\rm  for} \quad \phi_{cond}<\phi < \phi_{conf} \\
\nu \sqrt{d(d-1)}& {\rm  for}\quad \phi > \phi_{conf} ~.
\end{cases}
\eeq
It is straightforward to integrate now Eq.~\eqref{beta} for $\phi(a)$, and \eqref{integral}. The inner integrand is a smooth function of $z$ and can be estimated as a constant of order $z_s$.
The outer integrand in \eqref{integral} is highly peaked close to $\phi_{cond}$. Since the backreaction is still small at $\phi_{cond}$, it is a good approximation to take the the conformal coordinate as $z=1/a$, and one can straightforwardly integrate \eqref{integral}. Up to order-one factors, the result for small $\Delta_-$ is\footnote{The condensation and confinement scales in this model are related as 
${a_{conf}/ a_{cond}} \sim (\Delta_-)^{1/\widetilde\Delta_+}$.}
\beq\label{suppression}
m_\dil^2 \sim \big(\Delta_-\big)^{2-{d/\widetilde\Delta_+}}\;\Lambda_\IR^2 \,.
\eeq
Thus, $m_\dil$ is suppressed so long as $\Delta_-\ll1$ and the dimension  of the condensate is big enough, 
$$
\widetilde\Delta_+ > {d\over 2}~.
$$
Pictorially, the rise towards confinement needs to be fast enough. In hindsight, this seems quite natural. Unless the beta function has a sharp rise towards confinement (order $1$ values) there is no way that the beta function can possibly be small in the `IR'. 

Eq.~\eqref{suppression} correctly reproduces the suppression of
$m_\dil$ in models A and B. In model A, $\widetilde\Delta_+ =
\Delta_+=d-\Delta_-$, so now obtains \beq\label{mdilA} m_{\dil\,A}^2
\sim \Delta_-\; \Lambda_\IR^2 \,.  \eeq For model B, instead,
$\widetilde\Delta_+ \sim N\gg1$ and 
\beq
\label{mdilB} m_{\dil\,B}^2
\sim \Delta_-^2 \;\Lambda_\IR^2 \,.  
\eeq 
These behaviours are indeed observed in the numerical computation. Fig.~\ref{fig:formula} shows a comparison between the full result against formulas \eqref{integral1} and \eqref{integral}. The scalings \eqref{mdilA} and \eqref{mdilB} are
indeed reproduced.

Recall from the introduction that the CPR argument states that generically $m_\dil^2 \sim \beta_\IR$, even though additional suppressions can arise by tuning the IR brane potential. In the soft-wall realizations, the confinement plays the role of the hard IR brane and the condensate region controls the IR brane potential.

\section{Conclusions and discussion}
\label{sec:concl}

In this paper, we have studied confining deformations of CFTs driven by a single nearly-marginal scalar operator $\OO$. 
Generically, such deformations induce a nonzero condensate $\vev\OO$, and the spectrum of scalar excitations can contain a naturally light dilaton mode identifiable with the fluctuation of the condensate  $\vev\OO$. We found a mass-formula for the dilaton  involving an integral over the flow, 
\beq\label{integral2}
m_\dil^2\simeq\left[  
\int_{0}^{z_s} dz \; {1\over a^{d-1} \beta^{2}} \int_{z}^{z_s} dz'\,{a^{d-1} \beta^2}\right]^{-1}~,
\eeq
where $a$ is the RG scale, $\beta$ the beta-function, $z$ the bulk conformal coordinate that runs from 0 to  $z_s=\pi/\Lambda_\IR$, being $\Lambda_\IR$ the mass gap.
This formula can be thought of as a kind of average over the flow of the beta-function.  It has the remarkable property that whenever i) the $\beta$-function is small ({\em walking}) in the `IR' and ii) the rise in $\beta$ towards confinement ($\beta$ of order one) is fast enough, then the dilaton is light. 

In nearly marginal deformations,  ${\rm Dim}(\OO) = \Delta_+=d-\Delta_-$ with $\Delta_-\ll1$ and the deformation beta function can be estimated to be of order $\Delta_-$. The rise between walking  and confinement can be and is generically interpolated by the condensation of the operator $\OO$. It is illustrative to allow that ${\rm Dim}(\OO)$  changes along the flow. Let us call $\widetilde\Delta_+$ the effective dimension at the   condensation scale. Then \eqref{integral2} reduces to
\beq\label{suppression2}
m_\dil^2 \sim \big(\Delta_-\big)^{2-{d/\widetilde\Delta_+}}\;\Lambda_\IR^2~.
\eeq 
Hence, in order for the dilaton to be light  $\widetilde\Delta_+>d/2$, this gives a  quantitative idea of how fast the rise to confinement needs to be.
In the simplest models, with ${\rm Dim}(\OO)$ not changing along the flow ($\widetilde\Delta_+ = \Delta_+$), the suppression is linear $m_\dil^2 \sim \Delta_- \,\Lambda_\IR^2$. However, arranging for bigger $\widetilde\Delta_+$ gives a stronger suppression. This is exactly in line with the CPR proposal \cite{CPR}.

We have illustrated this by considering two families of holographic RG
flows with IR-walking and a confining soft-wall: models of `type-A',
which are dual to deformations of CFTs that are {\em generic} in that
they induce a condensate $\vev\OO$; and models of `type-B', which are
dual to deformations of CFTs that are {\em tuned} so that $\vev\OO=0$,
but which mock the condensation of an IR operator of large dimension
$\widetilde\Delta_+\gg d$. For the type-A models, we find $m_\dil^2
\sim \eepsilon \,\Lambda_\IR^2$, whereas for type-B models, $m_\dil^2
\sim \eepsilon^2 \,\Lambda_\IR^2$. The light dilaton is there in both
cases, regardless of whether $\vev\OO$ vanishes or not. This makes
perfect sense, on one hand because the dilaton mode is the fluctuation
of $\vev\OO$ and on the other because even if $\vev\OO=0$ the a
confining flow still breaks `spontaneously' scale invariance by
generating a mass gap.

Thus far, we have showed that there is a light state in these models. Its mass is suppressed by the order parameter of the breaking of conformal invariance (the beta function) evaluated at the condensation threshold (the largest of the scales that emerge in the IR). Clearly, then, this is a dilaton. 

Let us now comment on the technical naturalness of this construction. From the point of view of the gravity theory, we have certainly made an assumption on the form of the bulk potential $V(\phi)$. In the simplest (type-A) models, the potential for the bulk scalar $\phi$ is almost flat for small field $\phi\lesssim\phi_{conf}$, and exponential in $\phi$ for $\phi\gtrsim\phi_{conf}$. Such a potential might well be technically natural, because it interpolates between shapes that are protected by the shift symmetry $\phi \to \phi+$const. At $\phi\lesssim\phi_{conf}$, $V(\phi)$  is (approximately) invariant. At $\phi\gtrsim\phi_{conf}$, $V(\phi)$ scales by an overall constant under the shift. Of course, around  $\phi=\phi_{conf}$ there is no symmetry protecting  $V(\phi)$, and shift-symmetry breaking terms can permeate to small and large $\phi$ in the quantum theory. However, one expects that this communication is suppressed precisely in the limit that the transition is very sudden -- when the range $\Delta\phi$ with an unprotected potential is small $\Delta\phi/\phi_{conf}\ll1$. For instance, assume that the leading tail at small $\phi$ from the break at $\phi_{conf}$ is  $\phi^{2p}$ with integer $p$. This interaction certainly contributes radiatively  to {\em e.g.} the mass, but this starts at $p-1$ loops. Hence, for very large $p$ (very sudden transition) the communication is suppressed by many ($p-1$) loop factors\footnote{It is intriguing that the criterium for when $V(\phi)$ is natural is very similar to the criterium for when the dilaton is light -- they both require a fast transition.}. 
Interestingly enough, the form of the beta function (see Fig. \ref{fig:A}) seems quite insensitive to the details of the transition in $V(\phi)$ so long as it is fast enough. 
A more quantitative study of this protection is perhaps worth pursuing, but we take this as a very suggestive indication that this type of potentials are technically natural.

All of these results are perfectly in line with  the recent observation made by Contino, Pomarol and Rattazzi \cite{CPR} (that a naturally light dilaton can emerge provided the beta function `in the IR' is small). Their proposal is realized even in quite simple CFTs containing a single (nearly-marginal) scalar operator.

The mass-formula \eqref{integral2} has another obvious remarkable model-independent feature:  $m_\dil^2$ is explicitly positive, so this can be seen as a perturbative stability result. The dilaton turns out to be  never tachyonic, at least in confining models. This is not so surprising, though. The scalar-gravity system dual to a CFT with a scalar operator obeys positive energy theorems \cite{Amsel:2006uf,Faulkner:2010fh,Amsel:2011km}. This  allows to identify the regular RG flows as the ground state of the CFT deformation, and seemingly implies that the spectrum of glueballs  should not include any tachyons.

Let us also emphasize that Eq. \eqref{integral} has a wide applicability for confining flows -- it can be used as a diagnostic for the presence of a light dilaton. However, it is not necessarily applicable to other types of IR. For instance, in CFT deformations that flow to an IR CFT one would expect that the physical dilaton state acquires a finite decay width \cite{Dubovsky:2000am}. These cases are left for future investigation.

On the other hand, we have also elaborated on the method to compute the condensate $\vev\OO$ resulting from a (CFT-) deformation $\delta \LL = -\lambda\OO$. Holographically, the method is based on the fact that the holographic beta function satisfies a first-order differential equation, Eq. \eqref{betaEqn}. Supplemented with a regularity condition in the IR, this differential equation uniquely determines the condensate $\vev\OO$ once the holographic model (namely, the bulk potential $V(\phi)$) is specified. 
The origin of the differential equation Eq. \eqref{betaEqn}, obviously, lies on the fact that the scalar dual to  $\OO$ obeys a second order Klein Gordon equation. It seems, then, that locality in the holographic direction is what leads to the simple prescription to compute the condensates from a differential equation.

An immediate application of this method is that in theories (potentials) with a sharp transition to confinement, then i) a condensate generically develops and ii) the condensation scale is larger than the confinement scale. The ratio of scales goes like $a_{cond}/a_{conf}\sim (\Delta_-)^{-1/\widetilde\Delta_+}$. Here,  $\Delta_-$ is a proxy for the beta function at the condensation scale and $\widetilde\Delta_+$ for the effective dimension of the condensing operator. This suggests that the scale of the condensate should be larger than the confinement scale  even in theories with a moderate beta function at the condensation scale. Intriguingly enough, QCD seems to obey this rule: the scale of the gluon condensate (around $400-500 \,\textrm{MeV}$) is about twice of $\Lambda_{\textrm{QCD}}$ \cite{Shifman:1978bx}.

The method  is of course applicable to a very wide class of models -- beyond the confining flows considered here. In all cases one expects that for a given `model' (or bulk potential $V(\phi)$), generically $\vev\OO$ will not vanish. Also, if the flow experiences sudden changes (as in the flows discussed here), then one should expect condensate-dominated regions.

Concerning possible applications of the scenarios with a naturally light dilaton, the most obvious one -- adopting the light dilaton as the Higgs boson -- unfortunately seems to be already under some tension with LHC data \cite{Bellazzini:2012vz,Serra:2013kga}. However, given that in this type of models it is natural to have a considerable hierarchy of scales between the Higgs and the tower of resonances, it may be worthy to push this idea further. The other obvious target is the Cosmological Constant problem. Finding a similar natural mechanism to relax the Cosmological Constant (that is, to  by-pass Weinberg's no-go theorem \cite{Weinberg:1988cp}), seems far from straightforward \cite{Coradeschi:2013gda}. We hope to return to these issues in the future.

\paragraph*{Acknowledgements}

We are indebted with and deeply grateful to A. Pomarol, for suggesting to us the problem discussed in this paper, and for very valuable discussions during the whole duration of this project. We are also grateful to M.~Baggioli, B.~Bellazzini, A.~Pineda, J.~Serra, W.~Skiba and M.~Quir{\'o}s for discussions. 
We acknowledge support from Spanish MINECO under grants FPA 2011-25948,  Consolider-Ingenio 2010 Programme CPAN (CSD2007-00042) and Centro de Excelencia Severo Ochoa Programme grant SEV-2012-0234; and from Generalitat de Catalunya under grant 2009SGR894. E.M. is supported by a Juan de la Cierva fellowship (subprograma MICINN-JdC) and O.P. by a Ramon y Cajal fellowship (subprograma MICINN-RYC).

\end{document}